\documentclass[twocolumn,showpacs,showkeys,preprintnumbers,amsmath,amssymb,superscriptaddress,prl,aps]{revtex4-1}
\usepackage[dvips]{graphicx}
\graphicspath{{./}}

\usepackage{ifthen}
\usepackage{xcolor}
\usepackage{mathrsfs}
\usepackage{bm}
\usepackage{yfonts}
\usepackage{amsmath}
\usepackage{amssymb}
\usepackage{amsfonts}
\usepackage{upgreek}
\usepackage{wasysym}
\usepackage{mathbbol}

\usepackage{ulem}

\definecolor{DarkBlue}{rgb}{0.0,0,0.6}
\definecolor{DarkRed}{rgb}{0.6,0,0.0}

\DeclareMathAlphabet{\mathpzc}{OT1}{pzc}{m}{it}

\newcommand{\be}{\begin{equation}}
\newcommand{\ee}{\end{equation}}
\newcommand{\ba}{\begin{align}}
\newcommand{\ea}{\end{align}}
\newcommand{\bea}{\begin{eqnarray}}
\newcommand{\eea}{\end{eqnarray}}

\begin{document}
\title{Quantum Size Effects in the Atomistic Structure of Armchair-Nanoribbons}

\author{A. Dasgupta}
\affiliation{Institute of Nanotechnology, Karlsruhe Institute of Technology (KIT), D-76021 Karlsruhe, Germany}
\author{S. Bera}
\affiliation{Institute of Nanotechnology, Karlsruhe Institute of Technology (KIT), D-76021 Karlsruhe, Germany}
\affiliation{Institut f\"ur Theorie der Kondensierten Materie,
Karlsruhe Institute of Technology (KIT), D-76128 Karlsruhe, Germany}
\affiliation{DFG-Center for Functional Nanostructures (CFN), Karlsruhe Institute of Technology (KIT), Karlsruhe D-76131, Germany}
\altaffiliation[present address: ]{Institut N\'eel, CNRS
25 avenue des Martyrs, BP 166, 38042 Grenoble, France}
\author{F. Evers}
\affiliation{Institute of Nanotechnology, Karlsruhe Institute of Technology (KIT), D-76021 Karlsruhe, Germany}
\affiliation{Institut f\"ur Theorie der Kondensierten Materie,
Karlsruhe Institute of Technology (KIT), D-76128 Karlsruhe, Germany}
\affiliation{DFG-Center for Functional Nanostructures (CFN), Karlsruhe Institute of Technology (KIT), Karlsruhe D-76131, Germany}
\author{M.J. \surname{van Setten}}
\email{corresponding author: michiel.setten@kit.edu}
\affiliation{Institute of Nanotechnology, Karlsruhe Institute of Technology (KIT), D-76021 Karlsruhe, Germany}
\affiliation{DFG-Center for Functional Nanostructures (CFN), Karlsruhe Institute of Technology (KIT), Karlsruhe D-76131, Germany}
\date{\today}

\begin{abstract}
Quantum size effects in armchair graphene nano-ribbons (AGNR) with hydrogen termination are investigated via density functional theory (DFT) in Kohn-Sham formulation. ``Selection rules''  will be formulated, that allow to extract (approximately)  the electronic structure of the AGNR bands starting from the four graphene dispersion sheets. In analogy with the case of carbon nanotubes, a threefold periodicity of the excitation gap with  the ribbon width ($N$, number of carbon atoms per carbon slice) is predicted that is confirmed by ab initio results. While traditionally such a periodicity would be observed in electronic response experiments, the DFT analysis presented here shows that it can also be seen in the ribbon geometry: the length of a ribbon with $L$ slices approaches the limiting value for a very large width $1\ll N$ (keeping the aspect ratio small $N\ll L$) with $1/N$-oscillations that display the electronic selection rules. The oscillation amplitude is so strong, that the  asymptotic behavior is non-monotonous, i.e., wider ribbons exhibit a stronger elongation than more narrow ones.
\end{abstract}


\maketitle

The nonstandard electronic properties\cite{novoselov05nat,neto09rmp} along with improved fabrication techniques have moved graphene and its allotropes into the focus of frontier research in recent times.\cite{geim07natmat} The presence of an edge in graphene nano-ribbons (GNR) takes a significant influence on this electronic structure.\cite{han07prl,wakabayashi99prb,ritter09natmat,jiao09nat} In contrast to pure graphene, GNRs with a proper edge exhibit a finite bandgap potentially useful for device applications. Therefore, graphene nano structures with well defined orientation and edges have become a research field of their own.\cite{doessel11angew,jiao09nat,jia11nanosc,dubois09epjb,wassman09pssb,kosynkin09nat,wang10natchem,cai10nat,gosalbez-martinez11prb,vanin10prb,li10nanores,hamalainen11,bera10prb}

As is well known, zigzag-edged ribbons have two flat metallic bands near the Fermi energy possibly leading to magnetism.\cite{zarea09njp,wassmann11prl,palacios10sst} By contrast armchair ribbons (AGNR) are semiconducting with width-dependent band gaps and without affinity  to magnetic instabilities.\cite{blase10pssb,yu08ms,pisani07prb,barone06nl,rozhkov09prb,gunlycke08prb,yang07prl,son06prl,chen07,han07prl}
\begin{figure}[!tbp] \centering
\includegraphics[width=0.7\columnwidth]{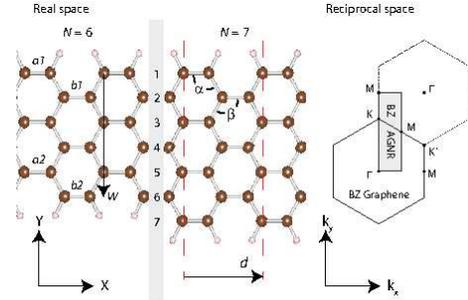}
\caption{(Color online) Left: Atomistic structure of armchair graphene nanoribbons (AGNR). They come in two species with odd and even number of carbon atoms $N$ in transverse direction. Right: The Brioulline zone of graphene appropriately parameterized for locating energy bands of the AGNR.}
\label{f1}
\end{figure}
In this letter we further investigate the electronic and atomistic structure of mono-hydrogenated AGNRs and show that interesting quantum effects arise, nevertheless. We formulate selection rules for the transverse momenta of the ribbon. They identify those lines in the (extended) Brioullin-zone of graphene that resemble the electronic structure of AGNRs of a given width $N$, the number of carbon atoms in transverse direction, see. Fig.~\ref{f1}. In this way we infer that a reasonable first approximation for all energy bands of an AGNR is encoded in a single selection rule. Similar to the case of a carbon-nanotube the selection rule predicts a three-fold periodicity of the bandgap in $N$. Because the same selection rule can be applied to all bands, one might suspect this periodicity to appear also in the atomic structure
of the ribbon. Indeed, this is what has been observed previously in the edge stress and energy for nonpassivated ribbons.\cite{huang09} Our detailed DFT-anatomy of AGNRs reveals however that the three-fold periodicity also appears in the atomic structure of hydrogen terminated AGNR: the longitudinal deformation of the unit cell of an
AGNR with respect to the bulk graphene value is described in leading harmonic approximation by a term $\sim\cos(2\pi N/3)/(N-1)$.

Our results imply that quantum size effects can be studied experimentally in AGNR by atomic structure determination,  namely by comparing the length of AGNRs (having the the same number of carbon slices, $L$) of neighboring width $N, N+1, \ldots$. This is in marked contrast to traditional approaches in meso- and nanoscopic structures that investigate quantum size effects near the band-gap by directly probing the electronic excitation spectrum, e.g. in transport measurements.

{\bf Method and model.} All DFT calculations presented in this letter are performed using a plane-wave basis set and the projector-augmented wave (PAW) method,\cite{paw,blo} as implemented in the Vienna {\it Ab initio} Simulation Package (VASP).\cite{vasp1,vasp2,vasp08} Based on a comparison of the graphene lattice-parameter obtained using several local, semi-local and meta-semi-local functionals (details are available in the supplemental material), the exchange-correlation functional of Perdew and Wang was chosen for the calculations.\cite{gga} For all geometries, the atomic structure is fully optimized using a conjugate gradient algorithm. A convergency criterium of 1~meV/\AA\ is used for the forces on the atoms.\footnote{Additional computational details are given in the supplemental material.}

\begin{figure}[!tbp] \centering
\includegraphics[width=0.8\columnwidth]{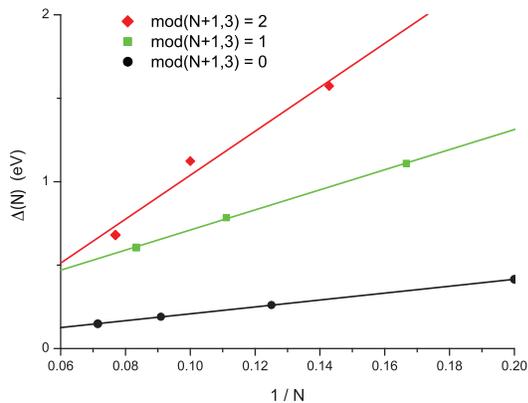}
\caption{(Color online) Bandgap $\Delta(N)$ of hydrogen terminated AGNRs as a function of $1/N$.}
\label{f4}
\end{figure}

{\bf Electronic structure of H terminated AGNR.} In Fig.~\ref{f4} the gap for electronic excitations of an AGNR is plotted over the inverse system width. The data splits into three sets exhibiting an oscillatory behavior.\cite{son06prl} For $N{+}1$ divisible by three the smallest gaps $\Delta(N)$ are seen and interpolate smoothly with $1/N$ into the bulk limit $\Delta{=}0$. For $N$ divisible by three the gaps are larger, also interpolating smoothly into $\Delta{=}0$. Largest gaps are encountered in the remaining case, $N{-}1$ divisible by three.

This overall phenomenology is reminiscent of the situation with carbon nanotubes (CNT)\cite{dresselhaussaito} and we will investigate it now closely. Zig-zag CNTs are similar to AGNR's with the H-termination replaced by periodic boundary conditions. The tube's electronic structure is understood by imposing selection rules for allowed transverse wavenumbers on graphene's band-structure in reciprocal ${\bf k}$-space. These selection rules reflect the cylindrical geometry. Three classes of armchair CNTs are thus obtained.\cite{dresselhaussaito} In a very similar way, also the electronic structure of AGNR can be understood as a cut through graphene's dispersion sheets along lines of selected transverse momenta $k_y$.

\begin{figure}[!tbp] \centering
\includegraphics[width=0.95\columnwidth]{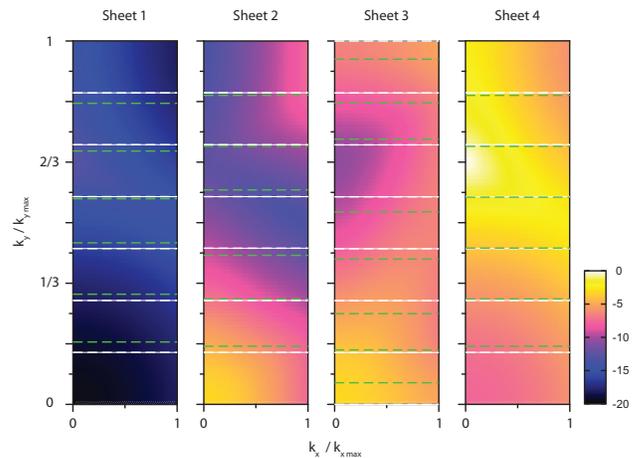}
\caption{(Color online) Energy bands for the $N=6$ AGNR (long dashed green lines) plotted into the graphene dispersion sheets in the ${\bf k}_x$, ${\bf k}_y$ plane. The dashed white lines indicate estimates based on the selection rule, see Eq.~(\ref{e2}).}
\label{f5}
\end{figure}

To find the appropriate selection rule for the AGNR's we map the (valence) bands $\epsilon(k_x)$ of the $N=5$ to $9$ ribbons onto the four (valence) dispersion sheets, $\epsilon(k_x,k_y)$, of pure graphene. The sheets are plotted as color maps in Fig.~\ref{f5}. For each sheet, $k_y$ values in graphene's Brioullin zone were selected such that the obtained set of $\epsilon_{k_y}(k_x)$ functions most closely resembles the band-structure of the AGNR. (The comparison of the functions for fitted $k_y$ to the actual ribbons bands is shown for all four sheets of the $N=5$ to $N=9$ AGNRs in Figs.~1 to~10 of the supplementary material.) The fitting process produces $N$ bands in sheets 1, 2 and 4. Due to the H-atoms at the edges the third sheet accommodates two more bands, $N+2$.

We infer from Fig.~\ref{f5} that in sheets 1, 2 and 4 the AGNR-bands are approximately equidistant with $\Delta k_y \approx k_\text{max}/(N+1)$ which motivates the following selection rule:
\be
\label{e2}
\{k_y\} = \bigcup_{i} \frac{i}{N + 1} k_\text{max}, \quad k_\text{max} = \frac{4\pi}{\sqrt{3}a_0},
\ee
where $i$ runs from 1 to $N$ for the sheets 1 to 3 and from 0 to $N+1$ for sheet 4 to also include the two $H$-bands.

\begin{figure}[!tbp] \centering
\includegraphics[width=0.8\columnwidth]{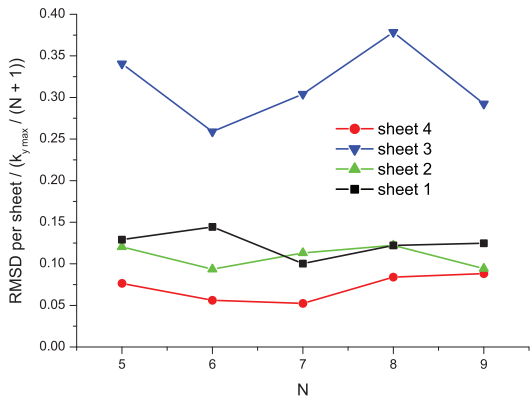}
\caption{(Color online) Root mean square deviation per sheet of the fitted $k_y$ values to the values predicted by the selection rule. The deviations in sheets 1, 2, and 4 remain smaller than 15\% of the line spacings independent of the ribbon width $N$. The bigger deviations in the fourth sheet result from the presence of the hydrogen edge.}
\label{aad}
\end{figure}

The AGNR-bands as predicted by the rules, Eq.~(\ref{e2}) are also indicated in Fig.~\ref{f5}; the root mean square deviation per sheet of the fitted values to the selection rule predictions are detailed in Fig.~\ref{aad}. We find a good match for the uppermost sheet 4, especially in the vicinity of the K-point. The overall discrepancy between the predicted and the true transverse momenta of the ribbon bands never exceeds 15\% of the interband spacing, except for the smallest ribbon widths. An exception to this rule is presented by the third sheet. Here, the presence of the $H$-atoms interferes and the selection rule gives only semi-quantitative information.

In analogy to the case of carbon nanotubes, also the proposed rules (\ref{e2}) allow for a qualitative understanding of the electronic structure. Indeed, for the AGNR with $\mod(N+1,3)=0$ the sheets are divided into a multiple of three equally wide sections. Hence, the $k_y=1/3$ line goes though the K-point (fourth sheet, Fig.~\ref{f5}) predicting a ribbon with an anomalously small bandgap.\cite{son06prl} After applying the selection rule to the other ribbons as well, we recover the values for the band gaps obtained by earlier tight binging calculations, with nearest neighbor hopping.\cite{son06prl,wang07,zhang09jpcc} In addition the selection also works for all the other sheets. The particularities of these sheets is not only encoded in the K-point behavior but also in the position of Van-Hove singularities. Since the positioning of the electronic levels with respect to these features also oscillates with period three, one expects that the oscillation behavior of the electronic structure also carries over to the atomic geometry.

{\bf Atomic Structure of H Terminated AGNR} We now investigate how the electronic structure translates into the atomistics of the AGNR, which we have also determined within our ab initio approach. For the cell geometry we adopt the nomenclature as depicted in Fig.~\ref{f1}. The extension of the unit cell in the $x,y$-directions is indicated by $d$ and $w$, respectively. For ease of comparison to bulk values we introduce the following dimensionless quantities: $\Delta d=\frac{d}{\sqrt{3}a_0}-1$, $\Delta w=\frac{w}{(N-1) (a_0/2)}-1$, and $\Delta A=(\Delta d+1)(\Delta w + 1)-1$, with $a_0$ the lattice parameter of bulk graphene calculated using the same functional and accuracy. All three become zero in the bulk limit, $N\to\infty$.

\begin{figure}[!tbp] \centering
\includegraphics[width=0.8\columnwidth]{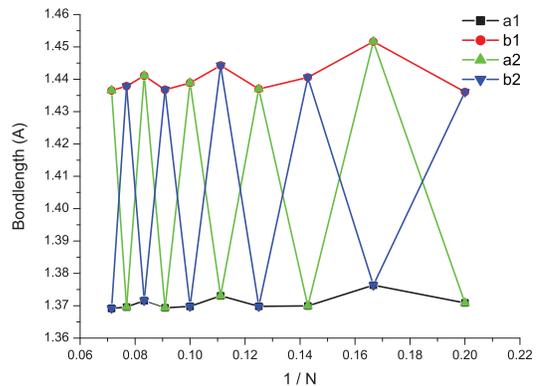}
\caption{(Color online) Evolution of the edge C-C bonds plotted against $1/N$. For the location of the bonds, a1, a2, b1, and b2, see Fig.~\ref{f1}. The growth of the sample width is in negative $y$-direction, so a2 and b2 oscillate, while a1 and b1 do not. The lower trace corresponds to the outer bonds, the upper trace to the (more) inner bonds.}
\label{f2}
\end{figure}

In the mono-hydrogen termination AGNR (H-AGNR) each carbon edge atom binds two other carbon atoms and one hydrogen atom. Considering the atomic structure of H-AGNR's this leads to a two-fold periodicity associated with $N$ even and $N$ odd, see Fig.~\ref{f1}. AGNRs of consecutive widths, $N=2n$ and $N=2n{+}1$, have different structural patterns; $2n{-}$AGNRs exhibit $n{-}1$ hexagons in all lateral sections
while $(2n{+}1){-}$AGNRs have a series of alternating $n$ and $n{-}1$ hexagons. The effect of this pattern can be visualized, by plotting the edge C--C bond-lengths, for the fully relaxed structures over the reciprocal of the ribbon width $1/N$, Fig.~\ref{f2}.
The way the edge exerts a pressure on the ribbon is, however nontrivial. Although the overall effect is elongating, we observe that the outer C--C bonds a1 are always shorter than the inner ones b1 (Fig.~\ref{f1}). Even more, the sum of a1 and a2 is shorter than two times the optimal bulk C--C. The difference survives in the large $N$ limit. The main cause of the elongation is hence not a change in bond lengths but the opening of the angles $\alpha$ and $\beta$.

\begin{figure}[!tbp] \centering
\includegraphics[width=0.8\columnwidth]{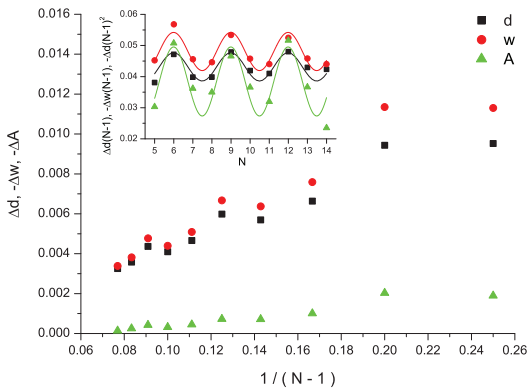}
\caption{(Color online) Relative deformation of the geometry of the unit-cell for hydrogen terminated graphene armchair nano-ribbons as a function of the inverse of the width $\frac{1}{N-1}$; Values for $N=5\ldots 14$ have been considered. The insets highlights the corresponding powerlaw dependencies on the ribbon width, which has the remarkable feature that the oscillations do not experience significant damping.}
\label{f6}
\end{figure}

The effect is also displayed in Fig.~\ref{f6} which now emphasizes healing of the geometry of the unit cell and allows for a simplified quantitative analysis:
$$
\Delta d =\kappa_d(N)/(N-1).
$$
The pre-factor is found to match the empirical form,
\be
\label{e4}
\kappa_d(N)\sim \kappa_d^{(0)}\left( 1 + \kappa_d^{(1)}\cos{(2\pi N/3)}\right),
\ee
with fitting parameters that can be read of the inset of Fig.~\ref{f6} ($\kappa_d^{(0)}\approx 0.04$ and $\kappa_d^{(1)}\approx 0.11$): it carries information about the strength of the edge-induced force per length, i.e., the chemical edge termination, and the elastic response of the AGNR which again incorporates the effective boundary conditions imposed on the ribbon's wavefunctions.

A striking aspect of Figs.~\ref{f2} and~\ref{f6} is the pronounced oscillation in $N$ with period three. It translates into the $\cos(2\pi N/3)$-term of Eq.~(\ref{e4}) that we take as evidence of the boundary conditions feeding back into the ribbons elasticity. By contrast, a two-fold periodicity originating from even/odd effects is largely suppressed.\footnote{The suppression of period two oscillations is understood in the following way. The most important contribution to the binding of sp$_2$-hybridized carbon atoms is made by in plane $\sigma-$bonds. Stretching at one edge of the AGNR faces a restoring force that originates from the inner $\sigma-$bonds. Most of this restoring force comes from innermost carbon atoms; only a small contribution is made by atoms of the opposing edge. For this reason the elastic deformation of the carbon-lattice responding to the H$_2$ induced stretch on one edge is hardly sensitive to the termination of the other edge.
\newline
As long as bond stretching is not too strong, we reside in the regime of linear elasticity. Then there is by definition no cross-talk between the stimulating stretching forces from either edge; the ribbons response is essentially a superposition of two independent stimuli, one from each edge, each one being largely insensitive to details of termination of the other edge. Hence, a two-fold periodicity is suppressed.}

It is a remarkable feature of Eq.~(\ref{e4}) that the oscillations do not experience significant damping (at least within the system sizes available to our numerics). In the absence of perturbations not included in our model such as ripples and other inhomogeneities, we expect that this statement should remain valid as long as the deviation between the true position of the AGNR-band and the prediction based on the selection rules, Eq.~(\ref{e2}), remains small as compared to $\Delta k_y$. Judging from Fig.~\ref{aad} this could be true at least in to the regime where the aspect ratio $N/L$ is still small.

Elastic material responses are usually (approximately) volume conserving. It is therefore reassuring to see that this is also the case in the present situation. Stretching the ribbon in longitudinal direction evokes a transverse contraction, see Fig.~\ref{f6}, which eliminates in the leading order the strain effect on the volume of the unit cell:
\be
\label{e5}
\Delta A  = \frac{\kappa_A(N)}{(N-1)^2}
\ee
where $\kappa_A(N)$ is of the form (\ref{e4}) with constants that can be read off the inset of Fig.~\ref{f6}:   $\kappa_A^{(0)}\approx 0.04, \kappa_A^{(1)}\approx 0.2$.

{\bf Conclusions.} We have studied quantum size effects in armchair graphene nano-ribbons (AGNR) with hydrogen termination using density functional theory (DFT) in Kohn-Sham formulation. By formulating ``selection rules'' that allow to extract (approximately) the electronic structure of the AGNR bands starting from the four graphene dispersion sheets, we have predicted a threefold periodicity of the ribbon's electronic structure in the ribbon width $N$ that was confirmed by ab initio results. We have also observed how this threefold electronic periodicity carries over into the atomic structure on AGNR.

Our results imply that quantum size effects can be studied experimentally in H-AGNR by atomic structure determination,  namely by comparing the length per carbon slice of AGNRs of neighboring widths $N, N+1, \ldots$. This is in marked contrast to traditional approaches in meso- and nanoscopic structures that investigate quantum size effects near the band-gap by directly probing the electronic excitation spectrum, e.g., in transport measurements. At a length of about 1850 carbon slices an $N=12$ ribbon will be a full chain of carbon atoms longer than an $N=11$ ribbon with the same number of chains, whereas the $N=13$ will be the same amount shorter. This difference is large enough that it could be detected experimentally. We expect that a termination other than hydrogen will only change amount of edge induced stress but not the mechanism underlying the elastic response. Hence, the values of the constants $\kappa$ will change but not the oscillating behavior.

\begin{acknowledgments}
Financial support by the Center for Functional Nano-structures (CFN), and cpu time allocation at the OPUS$^{\textrm{IB}}$ and hc3 clusters at the Karlsruhe Institute of Technology (KIT) Steinbuch Center for Computing (SCC) are gratefully acknowledged.
\end{acknowledgments}

\bibliography{D:/bibfiles/grapheneandribbon,../bibmine}

\end{document}